# Comparing Dawn, Hubble Space Telescope, and Ground-Based Interpretations of (4) Vesta

Vishnu Reddy
Max Planck Institute for Solar System Research, Katlenburg-Lindau, Germany
Department of Space Studies, University of North Dakota, Grand Forks, USA
Planetary Science Institute, Tucson, AZ 85719, USA
Email: reddy@psi.edu

Jian-Yang Li
Planetary Science Institute, Tucson, AZ 85719, USA

Lucille Le Corre
Planetary Science Institute, Tucson, AZ 85719, USA
Max Planck Institute for Solar System Research, Katlenburg-Lindau, Germany

Jennifer E. C. Scully
Institute of Geophysics and Planetary Physics, University of California Los Angeles, Los Angeles
California, USA

Robert Gaskell
Planetary Science Institute, Tucson, Arizona, USA

Christopher T. Russell
Institute of Geophysics and Planetary Physics, University of California Los Angeles, Los Angeles
California, USA

Ryan S. Park
Jet Propulsion Laboratory, California Institute of Technology, Pasadena, Califronia, USA

Andreas Nathues
Max Planck Institute for Solar System Research, Katlenburg-Lindau, Germany

Carol Raymond
Jet Propulsion Laboratory, California Institute of Technology, Pasadena, Califronia, USA

Michael J. Gaffey
Department of Space Studies, University of North Dakota, Grand Forks, USA

Holger Sierks
Max Planck Institute for Solar System Research, Katlenburg-Lindau, Germany

Kris J. Becker
Astrogeology Science Center, U.S. Geological Survey, Flagstaff, Arizona, USA

Lucy A. McFadden
NASA Goddard Spaceflight Center, Greenbelt, Maryland, USA

Pages: 51

Figures: 9
Tables: 5

**Proposed Running Head:** Dawn, HST, Ground-based Studies of Vesta

**Editorial correspondence to:**
Vishnu Reddy
59-495 Hoalike Road
Haleiwa
Hawaii 96712
 (808) 342-8932 (voice)
reddy@psi.edu

80  **Abstract**

81  Observations of asteroid (4) Vesta by NASA's Dawn spacecraft are interesting because
82  its surface has the largest range of albedo, color and composition of any other asteroid
83  visited by spacecraft to date. These hemispherical and rotational variations in surface
84  brightness and composition have been attributed to impact processes since Vesta's
85  formation. Prior to Dawn's arrival at Vesta, its surface properties were the focus of
86  intense telescopic investigations for nearly a hundred years. Ground-based photometric
87  and spectroscopic observations first revealed these variations followed later by those
88  using Hubble Space Telescope (HST). Here we compare interpretations of Vesta's
89  rotation period, pole, albedo, topographic, color, and compositional properties from
90  ground-based telescopes and HST with those from Dawn. Our goal is to provide ground
91  truth for prior interpretations and help identify the limits of ground-based studies of
92  asteroids in general. The improved rotational period measurement from Dawn is
93  0.222588652 day (Russell et al., 2012), and is consistent with the best ground-based
94  rotation period of 0.22258874 day (Drummond et al., 1998). The pole position for Vesta
95  determined by Dawn is 309.03º±0.01º, 42.23º ±0.01º and is within the uncertainties of
96  pole orientation determined by Earth-based measurements (Li et al., 2011: 305.8º±3.1º,
97  41.4º±1.5º). Similarly, the obliquity of Vesta is 27.46º based on the pole measurement
98  from Dawn and all previous pole measurements put the obliquity within 3º of this value.
99  Topography range from Dawn shape model is between -22.45 to +19.48 km relative to a
100 285 km x 285 km x 229 km ellipsoid. The HST range is slightly smaller (-12 km to +12
101 km relative to a 289 km x 280 km x 229 km ellipsoid) than Dawn, likely due to lower
102 spatial resolution of the former. We also present HST and Dawn albedo and color maps of Vesta in the Claudia (used by the Dawn team) and IAU coordinate systems. These maps serve to orient observers and identify compositional and albedo features from prior studies. We have linked several albedo features identified on HST maps to morphological features on Vesta using Dawn Framing Camera data. Rotational spectral variations observed from ground-based studies are also consistent with those observed by Dawn. While the interpretation of some of these features was tenuous from past data, the interpretations were reasonable given the limitations set by spatial resolution and our knowledge of Vesta and HED meteorites at that time. Our analysis shows that ground-based and HST observations are critical for our understanding of small bodies and provide valuable support for ongoing and future spacecraft missions.

# 1. Introduction

Vesta is one of the most frequently observed objects in the Main Asteroid Belt since its discovery by Heinrich Olbers in 1807. Ground-based color observations of Vesta as early as 1929 (Bobrovnikoff, 1929) revealed surface albedo/color variations that were attributed to composition. Taylor (1973) noted that Vesta's lightcurve changed depending on viewing geometry with northern hemisphere and equatorial views showing a single maximum. Degewij (1978) used polarimetry to verify that Vesta's lightcurve is dominated by albedo. The relationship between albedo and polarization and the overlap with the visual wavelength lightcurve suggested that the observed light curve was indeed controlled by albedo variation. Later photometric observations confirmed that hemispherical scale albedo variations dominate the lightcurve rather than shape (Drummond et al. 1988).

Disk-integrated visible wavelength (0.3 to 1.1 μm) spectral observations of Vesta (McCord et al. 1970) revealed a deep absorption band at 0.9 μm attributed to the mineral pyroxene. Overall spectral shape, and the presence of this pyroxene band, suggested a compositional link between Vesta and the howardites-eucrites-diogenites (HED) meteorites and suggested Vesta as a differentiated object. Rotationally resolved near-IR spectra of Vesta from NASA IRTF suggested that albedo variations might be linked to surface compositional heterogeneity (e.g., Gaffey, 1997; Vernazza et al. 2005; Reddy et al. 2010). Subsequent Hubble Space Telescope (HST) observations confirmed the affinity between compositional variations and albedo units thought to be surface morphological features (Thomas et al. 1997a; Binzel et al. 1997; Li et al. 2010).

The Dawn spacecraft entered orbit around Vesta in July 2011 to begin its yearlong mapping mission (Russell et al. 2012). During this period, the spacecraft mapped the surface of Vesta using its three instruments: Framing Cameras (FC), Visible and Infrared Mapping Spectrometer (VIR), and Gamma Ray and Neutron Detector GRaND). The Dawn Framing Cameras (FC) are a pair of identical 1024x1024 pixel imagers equipped with seven color filters (0.44-0.98 μm) and one panchromatic filter (Sierks et al. 2011) that imaged the surface of Vesta with an angular resolution of 93 μrad/pixel. Only FC2 was used during Vesta mapping phase with FC1 being a backup. Table 1 shows the list of filters along with their central wavelength and band pass. The FC mapped the surface at varying spatial resolution depending on the orbital phase. During approach, three rotational characterization (RC) phases imaged the entire visible surface at 9.07 km/pixel (RC1), 3.38 km/pixel (RC2), and ~487 meters/pixel (RC3, RC3B) (Table 2).

Dawn is the first mission to an asteroid that has been profusely studied by ground-based telescopes and HST over many decades. This wealth of knowledge not only helped us understand Vesta prior to arrival of Dawn, but also enabled us to verify the validity of these studies. With over 600,000 asteroids discovered so far, and the ever-increasing cost of robotic exploration of small bodies, sending a spacecraft to many of these asteroids is inconceivable. Dawn presents a unique opportunity to verify and validate ground-based observations of Vesta. In this work we aim to compare ground-based and HST data of Vesta with those from Dawn FC to verify interpretations of albedo units and rotational variations. We also provide maps in three coordinate systems: the original Olbers system based on Thomas et al. (1997a); the Claudia system based on Russell et al. (2012) and used by the Dawn science team; and the IAU coordinate system in which all the Dawn data will be archived on the Planetary Data System (PDS).

**2. Data Sets and Processing**

Three data sets were used in this paper: ground-based spectral data from Gaffey (1997) and Reddy et al. (2010); HST data from Thomas et al. (1997a), Binzel et al. (1997), and Li et al. (2010); and Dawn FC data from RC1, RC2 and RC3. The best resolution of HST data (38 km/pixel) is compared to the lowest resolution from Dawn during RC1 (9 km/pixel). Data reduction procedures for ground-based data are described in Gaffey (1997), Reddy et al. (2010); and for HST data processing in Thomas et al. (1997a), Binzel et al., (1997) and Li et al. (2010). A detailed description of Dawn FC data processing pipeline is presented in the supplementary materials section of Reddy et al. (2012a). Here we describe processing after the creation of photometrically and spectrally calibrated seven color global mosaics.

HST observations of Vesta at shorter wavelengths show higher contrast in albedo but this is below the wavelength range of Dawn FC filters. The Dawn albedo maps were created by using the 0.75-µm filter data which shows greatest albedo contrast among the FC filters and also has least amount of infield stray light residuals after correction (Reddy et al. 2012a). A global 0.75-µm-filter mosaic was extracted from the seven-color mosaic using IDL ENVI.

In addition to the albedo map, we created a band depth map, and a eucrite-diogenite (ED) ratio map. The band depth map (0.75/0.92 µm) is a single band color-coded map that helps to quantify the 0.9-µm-pyroxene band depth. The ED ratio is an interpretive scheme consisting of a single band rainbow color-coded map that uses the

195  ratio of 0.98 µm and 0.92 µm filters. Diogenite-rich areas are in red and show a deeper
196  0.90-µm pyroxene band whereas eucrite-rich areas are in blue (Reddy et al. 2012a). Due
197  to higher iron content in eucrites their 0.9-µm pyroxene band is shifted to longer
198  wavelength and so the eucrite ED ratio is ~1. In contrast diogenites have ED ratio >1 due
199  to their lower iron abundance (Reddy et al. 2012a).

## 3. Evolution of Coordinate Systems

201  The evolution of coordinate systems used on Vesta has historically depended on the
202  spatial resolution of data available at that time (Gaffey, 1997; Thomas et al. 1997b;
203  Russell et al. 2012). Typically, ground-based rotational spectral studies of asteroids (e.g.,
204  Reddy et al. 2010) used a lightcurve-based coordinate system where near-simultaneous
205  lightcurve observations are used to phase spectral observations (arbitrarily) with the
206  minima of the lightcurve becoming the prime meridian. Gaffey (1997) observed several
207  distinct compositional units on the surface of Vesta based on rotational spectral
208  observations. Creating a simple compositional map, Gaffey (1997) centered his prime
209  meridian on a compositional unit informally called "Leslie Formation," which was
210  interpreted as olivine-rich. This corresponds approximately to a weak inflection in
211  brightness around the lightcurve maxima of Vesta (0.75 rotational phase in Fig. 3 of
212  Gaffey 1997). It is important to note that the original maps published in Gaffey (1997)
213  had an error in latitude range. The correct latitude range of Gaffey (1997) maps is 90°N
214  to 60°S.

215  The general conventions for defining a body fixed coordinate system for Solar
216  System small bodies as adopted by the IAU/WGCCRE are summarized in Archinal et al.
217  (2010). Vesta is in simple rotation with slow precession (Asmar, pers. comm). For this case, its body fixed coordinate system follows the right-hand rule that defines a "positive pole" by its direction of angular momentum, and the longitude increases towards the direction of rotation. The positive pole of Vesta is above (or on the north of) the invariable plane of the Solar System (prograde rotation), similar to many planets, further simplifying the case without causing any confusion. Using the geodesy conventions, the positive pole of Vesta is thereby always referred to as the "North pole", and the direction of rotation is referred to as "East". All coordinate systems defined for Vesta in this work follow these conventions. In addition, when displaying maps, we adopted the convention to position the prime meridian at the center of the map, and longitude running between 0 and 360 (positive values). We adopted one single coordinate system and one single convention for all the maps and figures used here to avoid further confusion.

Historically, Gaffey (1997), Thomas et al. (1997a), and Binzel et al. (1997) maps were produced before the current conventions were adopted, and used a longitude system that increases towards the opposite direction of rotation, i.e., towards west, or a mixed east and west longitude to keep longitude values positive between 0º and 180º in the case of Thomas et al. But except for the longitude system, all those maps were displayed in a compatible orientation with the IAU conventions, i.e., 0-longitude center with north up. In addition, the map generated by Gaffey (1997) used incorrect range of latitude, while his north-south and east-west directions were compatible with the IAU conventions. The maps generated by Li et al. (2010) fully comply with the IAU conventions.

Thomas et al. (1997b) proposed a coordinate system for Vesta based on the rotational axis they derived from HST images and a new prime meridian centered on "the most prominent visible feature." This prominent feature was a 200 km wide low albedo unit, which was the largest and most distinct visible feature on Vesta, and was informally called "Olbers Regio" (Zellner et al. 1997) in honor of Vesta's discoverer Wilhelm Olbers. To calculate the location of the "Olbers" prime meridian for subsequent observations they developed a simple equation that used a rotation period of 0.2225887 day derived by Drummond et al. (1988). All subsequent publications (e.g., Binzel et al., 1997; Vernazza et al. 2005; Zellner et al. 2005; Carry et al. 2010; Li et al. 2010; Reddy et al., 2010, 2012b) till the arrival of Dawn at Vesta used the Thomas et al. (1997b) Olbers coordinate system with the Thomas et al. (1997b) pole measurement. Table 3 shows the values in pre-Dawn coordinate systems and in the Dawn coordinate system for the right ascension ($\alpha_0$) and declination ($\delta_0$) of the spin pole of Vesta and for the ephemeris position of the prime meridians. Binzel et al. (1997) updated the Vesta lithologic map from Gaffey (1997) into the Olbers coordinate system, and found that the prime meridian in Gaffey's maps was at ~255° east of Olbers.

Early in Dawn's encounter with Vesta the Dawn science team derived a coordinate system that was founded upon the new spacecraft-derived knowledge of the asteroid's surface. The system is defined by Claudia crater, which is a ~625 meter diameter crater located in Vesta's Oppia quadrangle (Russell et al. 2012). The location of the crater center is 1.6°S, 356.0°E in Claudia system coordinates. Vesta's prime meridian is 4° east of Claudia's position. Longitude increases to the east of the prime meridian. The Dawn team has undertaken all mapping and science investigations in the Claudia coordinate system.

The purpose of the Claudia coordinate system is to provide a convenient framework for representing and navigating the surface of Vesta, which is in accordance with the observations made by the Dawn spacecraft. Consistent with IAU best practices the Dawn Science team chose a small crater near the equator to anchor the coordinate system. The prime meridian was located for the convenience of the mappers, whose quadrangles are aligned with the prime meridian. It is traditional in mapping planetary surfaces to have the quadrangles all begin aligned with the prime meridian. The choice of the Dawn coordinates ensured that. No significant features were split by the prime meridian. Further, the prominent 'Snowman' craters, Marcia, Calpurnia and Minucia, were centrally located. It was believed to be safer for the users of the Dawn observations to work with a unique coordinate system that is clearly distinct from the early coordinate systems, which are based on various pole locations. These earlier coordinate systems (Li et al., 2011; Archinal et al., 2011; Seidelmann et al., 2007; Seidelmann et al., 2005; Seidelmann et al., 2002; Thomas et al., 1997) are nearer to the 180° meridian of the Claudia coordinate system. With the large difference between the Claudia system and the earlier systems, the use of the Claudia coordinate system can be easily visually verified. In practice, this choice of coordinate systems enabled those operating the spacecraft, and processing the data to spot errors by visual inspection. Coordinate system errors were not uncommon, but were found promptly with this device (S. Joy, personal comm., 2011).

    Claudia was chosen out of the innumerable small craters in the general location of the prime meridian, first because according to standard practices, it is close to the equator. Also it can be simply located by the use of a number of marker craters. Claudia is located roughly midway between Oppia and Gegania craters and to the north of Divalia Fossa. Claudia and eight larger craters form a cursive uppercase H (dotted lines in Fig. 1a), in which Claudia lies on the bar of the H. Near the midpoint of this bar are two

craters that are merged together and have a small crater in the center of their eastern rim (see Fig. 1b). Claudia is roughly 3.5 kilometers to the southeast of this small crater. Claudia was also chosen because it has a distinctive morphology, which makes it distinguishable from other similarly sized craters (see Fig. 1c). Claudia has a sharp, fresh rim; there is a ~170 meter diameter crater on the northeastern rim and two ~80 meter craters, along with many smaller craters, within Claudia's interior.

## 4. Comparison of Rotational Period and Pole Orientation

All early determinations of the rotational period of Vesta were performed through its rotational lightcurves (e.g., Cuffey, 1953; Groeneveld and Kuiper, 1954; Gehrels, 1967; Taylor, 1973; Chang and Change, 1962; Haupt, 1958; Magnusson, 1986). Although the rotational period of Vesta has been determined to an accuracy of $10^{-7}$ days, there was an ambiguity of whether Vesta's rotational period is 5.34 hr or 10.68 hr, rising from whether its lightcurve is single-peaked or double-peaked. Later observations using polarimetry, speckle interferometry, and radar favored the shorter period and showed that the surface of Vesta is variegated possibly due to compositional heterogeneity (Degewij et al., 1979; Drummond et al., 1988; Taylor et al., 1985). With the help from speckle interferometry and adaptive optics (AO), the surface of Vesta was resolved and the lightcurve of Vesta is shown to be single-peaked (Drummond et al., 1988). Using a large dataset of AO images, Drummond et al. (1998) determined a rotational period of 0.22258874 day with an uncertainty of 4 on the last decimal place (3.5 ms). This is the most accurate determination of Vesta's rotational period before Dawn's arrival at Vesta. The improved rotational period measurement from Dawn is 0.222588652 day with an uncertainty of 35 μs (Russell et al., 2012).

The pole orientation of Vesta was previously determined from ground-based and HST observations before Dawn arrived at Vesta with different datasets and techniques (Table 3). Thomas et al. (1997a) used HST images acquired in 1994 and 1996 with pixel sizes of 54 km and 38 km at Vesta, respectively, to construct a shape model of Vesta and determine its pole orientation based on surface albedo features and the limb profiles of the disk. They reported a pole orientation of (RA=301º, Dec=41º) with an uncertainty of 10º on both RA and Dec. Drummond and Christou (2008) fitted an ellipsoidal model to the limb of observed disk of Vesta from various observing and illumination conditions using ground-based, disk-resolved observations with speckle interferometry and adaptive optics techniques over the past 27 years. Their best-fit pole orientation was (306º, 38º) with an uncertainty of 7º. Li et al. (2011) combined all previous determinations of Vesta's pole orientation and their newly acquired images from HST in 2007 and 2010 at similar pixel size of previous HST images. They experimented with two methods, namely control point stereogrammetry and feature tracking, on all four HST datasets combined, and performed statistical studies of previous pole determination with ground-based data. The latest measurement of Vesta's pole orientation before Dawn measurements was (305.8º, 41.4º)±(3.1º, 1.5º) by Li et al (2011). While in orbit around Vesta, the Dawn spacecraft returned a substantial amount of imaging and tracking data, which have been used to precisely determine the pole orientation of Vesta as (309.03º, 42.23º) with an uncertainty of 0.01º (Russell et al., 2012). The obliquity of Vesta is 27.46º based on the pole measurement from Dawn. All previous pole measurements put the obliquity within 3º of this value (Table 3).

**5. Comparisons between HST and Dawn FC lightcurves**

HST observed Vesta four times in 1994 (Zellner et al., 1996), 1996 (Thomas et al., 1997b), 2007 (Li et al., 2010), and 2010 (Li et al., 2011), all with WFPC2 through the same set of filters; F439W, F673N, F953N, and F1042M. While the observing circumstances were all different, the 2007 observations have a sub-solar latitude of -5.3º, being the closest to the sub-solar latitude of Dawn data collected while approaching Vesta (south of -23º) among all HST observations. To compare the whole-disk lightcurves of Vesta collected by Dawn FC with those obtained by HST, we focus on HST 2007 observations and FC images collected on June 30, 2011 that cover one full rotation of Vesta. The observing geometries of both observations are listed in Table 4.

The shape of Vesta is close to a tri-axial ellipsoid, with the short axis aligned to its rotational axis. The difference between its intermediate axis and long axis is less than 3% (Thomas et al., 1997b; Russell et al., 2012). The rotational lightcurve of Vesta is single-peaked, and dominated by the albedo variations on its surface rather than by cross-sectional area variation or scattering geometry variation from global-scale topography. Due to moderate multiple scattering on Vesta (Li et al., 2012), the limb-darkening effect is stronger on Vesta than that on darker objects such as the Moon and C-type asteroids. Therefore albedo features near sub-solar point contribute slightly more to the shape of its lightcurve than those near the limb or terminator. For this reason, we plotted the brightness of Vesta with respect to sub-solar longitude rather than sub-observer longitude to compare the lightcurves obtained at different geometries. This is especially necessary for our case because the Dawn FC imaged the morning side of Vesta, while HST imaged the slightly afternoon side of Vesta, and the phase angles of both observations are quite different (Table 4).

356    Fig. 2 shows the lightcurves of Vesta measured from HST images (green) and
357    Dawn FC images (red and orange). At all three wavelengths, the overall agreement
358    between two observations is excellent in terms of the shape, amplitude, and phase of
359    lightcurves. The Dawn FC lightcurve is much smoother than the HST lightcurve,
360    presumably due to much higher signal-to-noise ratio in Dawn FC images measurement
361    where Vesta is resolved to 9.2 km/pix, compared to 38 km/pix in 2007 HST images. The
362    slight difference at sub-solar longitude from 60º to 330º could be due to viewing angle
363    difference. Dawn images have much more southern latitude than that of HST images, and
364    in mid northern latitude (30º-60º) the surface of Vesta appears to be brighter in this range
365    of longitude than in other longitudes.

366    Compared to the albedo maps (Fig. 4), the large bright and dark features in the
367    lightcurves (Fig. 2) all correspond to the large-scale albedo features. The lightcurve
368    minimum is located between longitude 270º and 330º, corresponding to the broad dark
369    area on Vesta observed in the same longitude. The lightcurve maximum is located
370    between longitude 60º to 120º, when the bright eastern hemisphere is visible. Compared
371    to the lightcurve minimum, the lightcurve peak appears to be slightly round-shaped or
372    plateaued. This is due to the large, relatively dark area near Oppia near ~100º longitude
373    surrounded by broad bright areas.

374 **6. Comparison of Vesta shape model from HST and Dawn**

375 The topography of Vesta was constructed using the technique of Gaskell et al. (2008)
376 from over 150000 small 99x99 pixel maplets, each representing the topography and
377 relative albedo of a small patch of Vesta's surface. Each maplet is located in an image by
378 illuminating its topography under the same viewing conditions as the image and performing a simple correlation. The center of the maplet represents a control point in image space, and due to the maplet's three-dimensional structure, its control point can be located in many images in a wide range of illuminations, resolutions and viewing geometries. This leads to an extremely precise solution for the location of the maplet's center. Similarly, the positions of many control points in a single image yield a precise determination of the Vesta-relative camera position and pointing at the image time (Gaskell, 2011). The maplets themselves are constructed by adjusting the slope and relative albedo at each of its pixels to minimize the residuals between the image brightness and illuminated maplet pixel brightness over a large number of images. The slopes are then integrated to produce the topography distribution within the maplet.

The ensemble of maplets can be combined to produce a global topography for Vesta's surface (Gaskell, 2012, private communication). Figure 3B shows this topography in a frame very close to the original Thomas et al. (1997a) coordinate system system. The heights range from -22.45 km (blue/violet) to +19.48 km (red) relative to a 285 km x 285 km x 229 km ellipsoid. Thomas et al. (1997a) were able to glean the topography of Vesta from images gathered by the Hubble Space Telescope. Their map, relative to a 289 km x 280 km x 229 km ellipsoid is shown in Figure 3A. There are striking similarities between the Thomas Hubble topography and the Dawn determined results. The Thomas range of heights was slightly smaller, but in general the agreement is very good and the differences can be explained in part by the dramatic difference in resolution between the two data sets.

Thomas et al. (1997a) note that the south pole (now called Rheasilvia) basin has a diameter of 460 km with an average depth below the rim of 13 km. The central peak within the basin is 13 km high compared to its deepest parts in HST map (Thomas et al. 1997a). In comparison, Dawn data show that the basin has a diameter of ~500±25 km and a depth of 19±6 km (Schenk et al. 2012). While these values are greater, they are consistent with HST model given the resolution of the data and the reference ellipsoids used. Thomas et al. (1997a) also note that a small section of the Rheasilvia rim is ~ 5 km higher than other parts. This region corresponds to Matronalia Rupes in Dawn data and rises higher than the rest of the basin rim as observed in HST data (Schenk et al. 2012).

## 7. Comparison of HST Observations and Dawn FC Data

*7.1 Global View*

Figure 4a shows the HST map of Vesta in 0.673-μm filter projected in the Thomas et al. (1997a) coordinate system based on observations from 1994, 1996 and 2007 oppositions at a resolution of ~50 km/pixel. The prime meridian here is defined through the dark feature informally named "Olbers" in Thomas et al. (1997a) located at ~15°N. The pole position for this data is $\alpha_0 = 301° \pm 5°$, $\delta_0 = 41° \pm 5°$. Figure 4b shows the Dawn FC map of Vesta in 0.75-μm filter from RC1 at a resolution of 9.06 km/pixel with the prime meridian similar to Thomas et al. (1997a) coordinate system in Fig. 4a. The pole position ($\alpha_0 = 309.03° \pm 0.01°$, $42.23° \pm 0.01°$) is the updated one from Russell et al. (2012) and hence features in the Dawn FC map are slightly rotated with respect to HST map. The Dawn data does not extend to northern latitudes in this map because of the spacecraft location with respect to Vesta during the RC1 phase. Fig. 4c is similar to Fig. 4b but the prime meridian is defined in the Claudia coordinate system (Russell et al. 2012) used by the Dawn science team in all their publications. The purpose of these three figures is to help ground-based observers orient themselves to features on Vesta from Dawn data.

*7.2 East-West Dichotomy*

Figure 4A-C also helps identify several hemispherical scale albedo features first observed in HST data by Thomas et al. (1997a). Despite the differences in pole position between Figures 4a and 4b, the Western hemisphere has an overall lower albedo than the Eastern hemisphere. Based on rotationally resolved near-IR spectra, Gaffey (1997) suggested that the dark hemisphere could be a howardite or polymict eucrite regolith that has been darkened by 'age-related darkening effect.' Binzel et al. (1997) was the first to detect this East-West dichotomy from HST images and interpreted the Western hemisphere as being "dominated by iron-rich and relatively calcium-rich pyroxene" similar to basaltic flows like eucrites. Zellner et al. (2005) suggested albedo features on Vesta could be impact craters/basins filled with dark material similar to lunar mare. Near-IR observations obtained by Vernazza et al. (2005) indicated that the Western hemisphere could be dominated by eucrite-type material. More recently, HST observations by Li et al. (2010) showed that this hemisphere is predominantly a large eucritic unit using the band ratio relationships associated with spectral characteristics of HED meteorites. Using ground-based spectral observations, Reddy et al. (2010) concluded that the Western hemisphere is dominated by howardite/polymict eucrite. Li et al. (2010) and Reddy et al. (2010) observations were also made at southerly sub-Earth latitude (-18.9°) compared to previous studies (16°) (Gaffey 1997 and Binzel et al. 1997).

Reddy et al. (2012a) studied the albedo and color variations on Vesta using Dawn FC color images. These albedo/color maps are in excellent agreement with HST albedo maps from Binzel et al. (1997) and Li et al. (2010) at hemispherical scale (Fig. 5 and 6). Interpreting the composition of the Western hemisphere, they concluded that it is dominated by a howardite and/or polymict eucrite, which is in agreement with ground-based (Gaffey, 1997; Vernazza et al. 2005) and HST observations (Binzel et al. 1997). However, Reddy et al. (2012a and 2012b) conclude that the cause of the lower albedo in the Western Hemisphere is not due to an age darkening effect (Binzel et al. 1997) or lunar mare type material filling impact craters/basins (Zellner et al. 2005) but due to the in fall of carbonaceous chondritic material. The presence of exogenous carbonaceous chondrite meteorite clasts in some howardite-eucrite-diogenite (HED) meteorites from Vesta is well documented (e.g., Buchanan et al., 1993; Zolensky et al., 1996). In HEDs, carbonaceous chondrite clasts generally make up to 5 vol. % (Zolensky et al., 1996), but on rare occasions can be 60 vol. % of howardites (Herrin et al. 2011). While some of these clasts have been heated and dehydrated during impact (~400°C), a majority of them are still hydrated (containing $H_2O^-$ and/or $OH^-$ bearing phases) (Zolensky et al., 1996). Carbonaceous chondrite clasts are the largest exogenic material observed in HED meteorites, but surprisingly they were not suggested as analogs for the dark hemisphere on Vesta by most ground-based and HST studies. Ground-based telescopic observations of Vesta in the mid-IR indicated the possible presence of a 3-µm absorption feature (Hasegawa et al. 2003). Hasegawa et al. (2003) noting the presence of a 3-µm absorption suggested contamination from impacting carbonaceous chondrites as possible cause of this feature.

Rivkin et al. (2006) observations did not find the same feature reported in Hasegawa et al. however, the 2006 observations were at a different viewing geometry than 2003 and did not extend as far North as 2003, and variability of spatial extent of the 3-µm feature was offered as a possible explanation. A weak absorption band at 2.8 µm attributed to OH adsorption is reported with variable expression in the spatial domain (DeSanctis et al., 2012b). Evidence of hydrogen enrichment in the same region has been found by Dawn (Prettyman et al. 2012) and confirms the ground-based evidence for a water-related feature near the vibrational bands of hydrated species.

Gaffey (1997) and Binzel et al. (1997) interpreted the higher albedo Eastern hemisphere as impact excavated plutonic material similar to diogenites. They also identified several compositionally distinct units using band parameter systematics and concluded that several Eastern hemisphere units have substantial olivine component (more discussion on olivine in Section 6.3). Vernazza et al. (2005) and Carry et al. (2010) also interpreted spectral variations in near-IR to indicate a substantial diogenite in the Eastern hemisphere of Vesta using ground-based spectral and adaptive optics data. Li et al. (2010) and Reddy et al. (2010) concluded there might be a large diogenite unit straddling the boundary between the two hemispheres.

Reddy et al. (2012a) mapped the abundance and distribution of diogenite-rich material on Vesta using Dawn FC color images and color indicators of HED components. Albedo (0.75-µm filter), pyroxene band depth (0.75/0.92-µm filter ratio) and eucrite-diogenite (ED) ratio (0.98/0.92-µm filter ratio) maps of Vesta (Fig. 6a-c) from Dawn FC confirm the presence of diogenite-rich material in the southern hemisphere. The ED ratio map shows diogenite-rich areas in red and eucrite-rich areas in blue (Fig. 6c). The distribution of diogenite is predominantly restricted to the southern hemisphere except for a small region that stretches north between 120°-180° W longitude that Reddy et al. (2012a) interpreted as ejecta from the Rheasilvia basin.

*7.3 North-South Dichotomy*

494    Thomas et al. (1997a) studied the southern hemisphere of Vesta using HST. They noted
495    that the band depth and width of the 0.9-μm pyroxene absorption increased with
496    "excavation depth, consistent with exposure of a higher calcium content and coarser
497    grained pyroxene-rich plutonic assemblage within the crust of Vesta or the exposure of
498    olivine present within the upper mantle or both." Reddy et al. (2010) observed the
499    southern hemisphere of Vesta using the IRTF and concluded from near-IR absorption
500    band position that Vesta's southern hemisphere is more diogenitic than the northern
501    hemisphere. These conclusions were consistent with Li et al. (2010) HST observations of
502    the Southern hemisphere of Vesta.
503    The ED ratio map of Vesta produced from Dawn data (Fig. 6c) shows more
504    diogenite-rich material (red) in the southern hemisphere centered around the Rheasilvia
505    basin and bluer eucritic areas in the north confirming ground-based (Gaffey 1997, Reddy
506    et al. 2010) and HST observations (Thomas et al. 1997; Li et al. 2010) of Vesta. There is
507    also a distinct correlation between the albedo (Fig. 6a), band depth map (Fig. 6b), and ED
508    map (Fig. 6c). Brighter areas in the albedo map correspond to areas with deeper band
509    depth and diogenite-rich areas and conversely darker areas correspond to those with
510    weaker band depth and more eucrite-rich areas.
511    Reddy et al. (2012a) explored the cause of deeper band depth in diogenite-rich
512    areas and concluded that this could be a particle size effect as suggested by Thomas et al.
513    (1997a). Based on meteoritic evidence, the average grain size of diogenites in howardite
514    breccias ranges from 500 μm to 1.5 millimeters whereas the average grain size of eucritic
515    material is ≤70 μm (Beck and McSween, 2011). This difference could be due to a
516    combination of surface composition and regolith processes (Chamberlain et al. 2007).

517  Diogenites cooled at depth with coarser (pyroxene) grains compared to eucrites, which
518  are typically fine-grained basalts (surface flows). The larger grain size for diogenites
519  could also be explained by shorter surface exposure ages (fresher impacts) compared to
520  eucrites that might have experienced longer exposures to regolith processes on the
521  surface (impact gardening) (Reddy et al. 2012a). The latter would indicate Rheasilvia
522  basin forming more recently excavating diogenite rich material.

523  *7.4 Cross correlation between HST and Dawn FC Albedo Maps*

524  Binzel et al. (1997) identified 19 different albedo and color units on Vesta using HST
525  observations. These alphabetically ordered units were selected for their spectral diversity
526  that included albedo, band depth, and width (ED). Following similar methods, Li et al.
527  (2010) expanded this list to include 15 new features (numerically ordered) in the southern
528  hemisphere. Of these 34 features identified in HST images, a few are prominently visible
529  in albedo maps while others stand out in band depth and ED ratio maps. Figure 7A shows
530  those units in the albedo map of Vesta from HST in 0.673-µm filter and Fig. 7b shows
531  the same units as identified in the albedo map from Dawn FC in 0.75-µm filter. Both
532  maps are in Thomas et al. (1997a) coordinate system.

533  Albedo features from the HST map have been identified and labeled on the Dawn
534  FC map by correlating albedo patterns visually and confirming them based on the
535  description of their spectral properties from Binzel et al. (1997) and Li et al. (2010). The
536  most prominent albedo features that are easily identifiable in both maps include low
537  albedo features such as "Olbers," features #8, #9, A, and B within the dark unit, features
538  Q and #15. High albedo units that are relatively easy to identify in both maps include
539  features #4, #15, #13, #11, and Z. Given the differences in spatial resolution (50 km/pixel

540 vs. 9 km/pixel) and pole orientation, some diffuse features in the HST map (Fig. 7a)
541 cannot be confirmed in the Dawn FC map shown in Fig. 7b. A complete list of HST units
542 and their corresponding surface features with IAU approved names from Dawn FC
543 images is in Table 5.

544 *7.5 Interpretations of specific features*

545 Ground-based rotationally resolved spectroscopy of Vesta led to the identification of
546 specific compositional units on Vesta (Gaffey, 1997). By correlating variations in
547 spectral parameters (Band Centers and Band Area Ratio or BAR) with rotational
548 lightcurve of Vesta, the location of distinct mineralogical units could be constrained
549 longitudinally and helped create the first maps (Gaffey, 1997). Subsequent disk-resolved
550 color filter observations by HST (Thomas et al. 1997) led to the identification of albedo
551 features associated with these units (Binzel et al. 1997) in the northern hemisphere. Li et
552 al. (2010) identified additional compositional units in the southern hemisphere of Vesta
553 from HST images and Reddy et al. (2010) linked these to specific mineralogical units
554 based on ground-based spectra.

555    Thomas et al. (1997a) observed three distinct regions that had showed variations
556 in band depth and band width. These include the Rheasilvia basin (discussed in section
557 6.1), Matronalia Rupes on the rim of the South Pole crater, and a potential crater.
558 Matronalia Rupes feature on Dawn FC maps corresponds to feature #7 (Table 5) on HST
559 maps. This 20-km high scarp (Schenk et al. 2012) stretches for nearly 200 km around the
560 Rheasilvia basin. Thomas et al. (1997a) noted a reverse compositional trend in
561 Matronalia Rupes "consistent with composition of excavated material that has been
562 overturned." Dawn FC color images show Matronalia Rupes is dominated by diogenite-rich material (Reddy et al. 2012a) excavated during the formation of the Rheasilvia basin consistent with Thomas et al. (1997a) interpretation. Li et al. (2010) also identified several features (#4, #5, #6) that are interpreted as diogenite-rich areas "located on high rims of the crater." All these units are located along the rim of the Rheasilvia basin or within the diogenite-rich ejecta blanket that spreads north between 180° and 120°W longitude in Fig 6A-C. This ejecta blanket and diogenite units have also been observed in ground-based compositional map from Gaffey (1997) and Reddy et al. (2010).

Gaffey (1997) noted a decrease in ratio of 2-µm pyroxene band (Band II) area to that of 1-µm pyroxene band (Band I) area from ground based spectral observations as Vesta rotated. A well established cause for decrease in BAR is increase in the abundance of olivine in a mixture of olivine and orthopyroxene (Cloutis et al. 1986). Based on this interpretation Gaffey (1997) suggested the presence of a large olivine-rich unit on the surface of Vesta located between 60-120° E longitude (Fig. 8-9). This olivine-rich unit geographically corresponds to feature #15 from HST map of Li et al. (2010) (Fig. 7A-B). This feature has the reddest spectra color over the entire surface of Vesta observed by the HST (Li et al. 2010) with band depth shallower than the background surface.

In Dawn FC color maps (Fig. 6A-C) this feature corresponds to the Oppia crater and its ejecta blanket. Reddy et al. (2012a) observed that Oppia has the reddest visible spectra slope (0.75/0.44 µm) on Vesta consistent with Li et al. (2010) HST observations. The band depth (0.75/0.92 µm) of Oppia region is also shallower than the average surface of Vesta (Reddy et al. 2012a). Due to low albedo, red slope and weaker pyroxene bands, Li et al. (2010) interpreted this region to be space weathered (Fig. 8). Investigating the composition of Oppia region, Le Corre et al. (2012) estimated the BAR as 1.57, which is much lower than the average BAR (2.74) for Vestan surface (Gaffey, 1997). Le Corre et al. (2013) attributed this low BAR to impact melt or excavation of cumulate eucrite layer by the Oppia impact. Hence the low BAR for Oppia region, which was interpreted as olivine by Gaffey (1997), might not be interpreted so today. The best way to determine the presence of olivine on Vesta's surface is through the interpretation of higher resolution data from Dawn. We also rule out selective lunar-style space weathering as the cause of the red visible slope and lower BAR observed in color and spectral data of Oppia (Pieters et al. 2012).

As noted earlier, Zellner et al. (1997) unofficially named a prominent low albedo feature on Vesta as "Olbers Regio" in honor of its discoverer Wilhelm Olbers. This feature also defined the location of the prime meridian in HST maps. Binzel et al. (1997) interpreted this low albedo feature as "remnants of Vesta's ancient basaltic crust." They also noted that the pyroxene band depth of the Olbers feature was relatively shallow and narrow suggesting the predominance of a single pyroxene. Invoking lunar-style space weathering to explain the low albedo and shallow pyroxene absorption band of Olbers, Binzel et al. (1997) concluded that Olbers is remnant ancient crust.

Dawn FC observations of Vesta show that the Olbers region in the HST map corresponds to the dark area East of Marcia, Calpurnia, and Minucia craters. Consistent with HST observations by Binzel et al. (1997) this region has low albedo and weaker pyroxene band compared to Vesta global average. Prettyman et al. (2012) noted an enhancement in hydrogen signature corresponding to Olbers region based on GRaND observations of Vesta and attributed this to the presence of $H_2O/OH$ in carbonaceous chondrite xenoliths. Although our study of this region is still ongoing, the ejecta distribution around Marcia suggests that it is younger than low albedo region east of it. The low albedo and weaker pyroxene bands of Olbers region are probably due to the mixing of carbonaceous chondrite impactor material and ancient howardite regolith.

One of the brightest albedo spots in HST observations of Vesta is feature #13 from Li et al. (2010). They observed that this feature has an extremely blue spectral slope and is located on the steepest slope(s) on Vesta. The depth of the pyroxene absorption band of this feature is indistinguishable from the background surface in HST observations. In ground based observations of Vesta (Binzel et al. 1997; Gaffey, 1997) this feature corresponds to a diogenite-rich unit located ~45° S latitude. In Dawn FC images, this feature corresponds to ejecta around the Antonia crater, a 17.4–km diameter impact feature within the Rheasilvia basin. The Framing Camera shows that Antonia crater is indeed located on a topographically steep slope on Vesta consistent with HST observations by Li et al. (2010). ED ratio map of Vesta (Fig. 6b-c) showed that the region around Antonia crater is dominated by diogenite-rich material consistent with ground-based observations by Gaffey, (1997).

Of the six geologic units identified by Li et al. (2010) in HST images (Fig. 9), area IV (60° E longitude) showed a deep pyroxene band, neutral red slope and was interpreted as diogenite rich. In Dawn FC data (ED map), this region corresponds to 30-km diameter Numisia crater, which lies SE of Olbers region. Numisia has been interpreted as diogenite-rich consistent with HST observations.

*7.6 Space Weathering*

Space weathering consists of processes by which the surface optical properties are modified when exposed to space environment in the absence of an atmosphere. On the Moon the primary effect of space weathering is decrease in optical albedo, increase in spectral slope (reddening), and a decrease in absorption band depth (Pieters et al. 2000; Taylor et al. 2001). Nano phase iron created during space weathering coats the lunar soil grains causing these observed effects (Loeffler et al. 2008). In contrast to the Moon, asteroids visited by spacecraft prior to the arrival of Dawn at Vesta showed distinctly different effects of space weathering (Gaffey, 2010). On (243) Ida, the band depth decreased and the spectral slope increased, while the albedo remained nearly constant with increased level of space weathering (Veverka et al. 1996). On (433) Eros, the overall albedo decreased but the spectral slope and band depth remained nearly constant with increased level of space weathering (Murchie et al. 2002).

Space weathering on Vesta has been a source for intense scrutiny over the last two decades because of its similar basaltic composition as that of the Moon. Detection of large-scale albedo variations on Vesta using HST suggested that some of them might be ancient 'space weathered' units (Binzel et al. 1997). Hiroi and Pieters (1998) conducted simulated space weathering experiments on HED meteorites and compared them to Vesta and Vestoids. Vestoids are small asteroids (~10 km) that have spectral properties similar to Vesta and are part of the Vesta dynamical family. They found that Vesta and some Vestoid spectra showed effects similar to lunar style space weathering, but some Vestoid spectra did not follow the "HED-lunar space weathering trend." They concluded that these Vestoids could be from a different parent body. Addressing the cause of "bluer" visible spectra of Vesta, Vernazza et al. (2006) invoked the possibility of a magnetic field around Vesta that could be protecting the surface against lunar-style space weathering effects. Meteoritical evidence for space weathering products in Vestan regolith has been limited. Noble et al. (2011) searched for space weathering products (agglutinates and nano phase iron bearing rims) in howardite Kapoeta and found several "melt products, including spherules and agglutinates and possible nano phase bearing rim." They concluded that the presence of these products would suggest that lunar style space weathering is active on asteroids like Vesta.

Analyzing Dawn FC and VIR spectral data, Pieters et al. (2012) noted that Vesta's surface shows effects of space weathering but it is unlike that seen on the Moon. They also found no evidence for the accumulation of nano phase iron, which is the principle cause of lunar space weathering effects, on Vesta's regolith particles. Gradual fading of freshly exposed material seen on Vesta has been attributed to localized mixing of material with diverse spectral and albedo properties (Pieters et al. 2012, McCord et al. 2012, Li et al. 2012). McCord et al. (2012) have shown that impact mixing of bright pristine material (Li et al. 2012) with darker exogenous carbonaceous chondrite material (Reddy et al. 2012) gives rise to the background gray material. In light of this new information from Dawn, it appears that general application of lunar style space weathering to all asteroids is invalid. This result is also consistent with space weathering trends observed on other asteroids by spacecraft (Ida and Eros) that suggest unique styles of space weathering operating on each of these objects (Gaffey, 2010).

## 8. Future Earth-Based Observations

Studies of Vesta by ground-based and orbiting telescopes helped the Dawn mission to be better prepared to explore Vesta. While the results from the mission has confirmed many pre-Dawn characterizations of Vesta, a myriad of surprises were revealed, which could only be possible with an orbital mission. Based on the comparison of ground-based and HST data of Vesta with Dawn's Framing Camera data, we can conclude that ground based observations have been extremely accurate at constraining rotation period, surface composition using reflectance spectroscopy, and detecting/mapping rotational spectral variations on small bodies. HST observations have also been successful in accurately identifying albedo and color features, surface composition of these features and constraining the shape, pole orientation and topography. The success of these precursor studies can be attributed to two factors: a) availability of high quality ground-based and HST data, and b) interpretations backed by laboratory studies of HED meteorites that were delivered to Earth from Vesta.

However, ground-based observations and their interpretations have severe limitations in accessing the prevalence and magnitude of space weathering on individual small bodies. As noted earlier, space weathering on Vesta (Pieters et al. 2012) and other small bodies visited by spacecraft has been shown to be unlike that on the Moon (Gaffey, 2010). On Vesta, alternate processes such as presence of exogenous opaques (Reddy et al. 2012b; McCord et al. 2012), observing geometry/phase angle (Reddy et al. 2012c; Sanchez et al. 2012) and particle size (Reddy et al. 2012a; Pieters et al. 2012) have been shown to be the cause of traditional lunar-style space weathering effects like low albedo, red spectral slope, and weaker absorption bands. The effects of multiple impacts from large and small projectiles control the distribution and mixing of materials in Vesta's regolith. Future interpretations of ground-based spectra of small bodies must take into account these findings from Dawn and other missions before interpreting red spectral slope in ground based spectra of asteroids as an indicator of space weathering.

## 9. Summary and Conclusions

Ground-based and HST observations of Vesta provided valuable insight into the asteroid's composition, surface albedo and topography prior to the arrival of NASA' Dawn spacecraft. Our comparative study reveals the following:

- Rotation period of Vesta as determined by the Dawn spacecraft (Russell et al., 2012) is 0.222588652 days with an uncertainty of 35 µs. This is consistent with Drummond et al. (1998) rotational period of 0.22258874 day, the most accurate determination of Vesta's rotational period before Dawn's arrival at Vesta.

- Dawn spacecraft has precisely determined pole orientation of Vesta to be (309.03º, 42.23º) with an uncertainty of 0.01º (Russell et al., 2012). This is a substantial improvement over the best ground/HST based pole position (305.8º, 41.4º)±(3.1º, 1.5º) by Li et al. (2010).

- While the Dawn FC lightcurve is much smoother than the HST lightcurve due to much higher signal-to-noise ratio, the overall agreement between the two is excellent in terms of the shape, amplitude, and phase of lightcurves.

- HST topographic map from Thomas, et al (1997) is consistent with Dawn shape model. There are striking similarities between HST topography and the Dawn determined results. HST range of heights was slightly smaller (-12 km to +12 km) than Dawn (-22.45 km to +19.48 km), likely due to much lower spatial resolution of the former, but in general the agreement is very good.

- Global albedo/color maps of Vesta using HST (Binzel et al. 1997; Li et al. 2010) are in excellent agreement with Dawn FC maps (Reddy et al. 2012). East-West hemispherical dichotomy observed in ground-based (e.g., Gaffey 1997) and HST images (Thomas et al. 1997a) has been confirmed by Dawn FC color images.

- Ground-based telescopic observations indicated the possible presence of a 3-μm absorption feature. Hasegawa et al. (2003) and Rivkin et al. (2006) suggested contamination from impacting carbonaceous chondrites as possible cause of this feature. Dawn observations of Vesta (De Sanctis et al. 2012b; Prettyman et al. 2012; McCord et al. 2012; Reddy et al. 2012b) confirm these ground-based observations.

- Predominance of diogenite-rich material in the higher albedo Eastern hemisphere observed by ground-based observers (Gaffey 1997; Binzel et al. 1997; Vernazza et al. 2005; Carry et al. 2010; Li et al. 2010) has been confirmed by Dawn (Reddy et al. 2012a).

- A majority of albedo and color units observed in HST images of Vesta (Binzel et al. 1997; Li et al. 2010) have been confirmed by Dawn FC observations.

- Dawn data supports an alternative interpretation of "olivine-rich" unit observed by Gaffey (1997). The low BAR for Oppia region, which was interpreted as olivine by Gaffey (1997), would not be so interpreted based on Dawn data (Le Corre et al. (2011).

- Ground-based observations have severe limitations in identifying space weathering and constraining its magnitude on individual small bodies. Dawn observations have demonstrated that lunar-style space weathering does not operate on Vesta, despite both objects having a basaltic surface composition (Pieters et al. 2012).

- Precursor studies of Vesta have been very successful in accurately characterizing its physical properties. This can be attributed to two factors: a) availability of high quality of ground-based and HST data, and b) interpretations backed by laboratory spectra and studies of samples of HED meteorites

**Acknowledgement**

We thank the Dawn team for the development, cruise, orbital insertion, and operations of the Dawn spacecraft at Vesta. The Framing Camera project is financially supported by the Max Planck Society and the German Space Agency, DLR. We also thank NASA's Dawn at Vesta Participating Scientist Program for funding the research. A portion of this work was performed at the Jet Propulsion Laboratory, California Institute of Technology, under contract with NASA. Dawn data is archived with the NASA Planetary Data System. VR would like to thank Juan Andreas Sanchez and Guneshwar Singh Thangjam for their help in improving the manuscript.

**Table 1.** List of FC filters (except clear filter) with their respective band passes width and peak.

| Filter name | Wavelength center (μm) | FWHM (μm) |
|---|---|---|
| F8 | 0.438 | 0.040 |
| F2 | 0.555 | 0.043 |
| F7 | 0.653 | 0.042 |
| F3 | 0.749 | 0.044 |
| F6 | 0.829 | 0.036 |
| F4 | 0.917 | 0.045 |
| F5 | 0.965 | 0.086 |

**Table 2.** Observational circumstances for Dawn data collected at Vesta.

| Orbital Phase | Best Resolution (meters/pixel) | Sub-Spacecraft Latitude | Distance To Vesta (km) |
|---|---|---|---|
| RC1* | 9067 | -32° | 100,000 |
| RC2 | 3382 | -54° | 37,000 |
| RC3 | 487 | -25° | 5,200 |
| RC3B | 487 | -25° | 5,200 |
| Survey | 252 | 50° to -90° | 2,700 |
| HAMO** (1) | 61 | 66° to -87° | 660 to 730 |
| LAMO*** | 16 | 85° to -90° | 190 to 240 |
| HAMO (2) | 60 | 85° to -85° | 640 to 730 |

*Rotational Characterization
**High Altitude Mapping Orbit
***Low Altitude Mapping Orbit

Table 3. Values for right ascension, declination of the spin pole, ephemeris position of the prime meridian, obliquity, right ascension of vernal equinox, longitude of Olbers Regio, longitude of Claudia and long axis longitude in pre-Dawn coordinate systems (I to VI) and Claudia coordinate system (VII). Obliquity is the angular distance between the orbital and rotational poles and vernal equinox is the descending intersection of Vesta's orbital and its equatorial planes. Note that the IIb values were later used by Seidelmann et al., 2001, Seidelmann et al., 2005 and Seidelmann et al., 2007.

| | Source | $\alpha_0$ (RA of Spin Pole) | $\delta_0$ (Dec. of Spin Pole) | W (Ephemeris Position of Prime Meridian) | Obliquity | Right Ascension of Vernal Equinox | Longitude of Olbers Regio | Longitude of Claudia | Long Axis Longitude |
|---|---|---|---|---|---|---|---|---|---|
| I | Drummond et al., 1988 | 335° ± 4° | 41° ± 4° | 299° + 1617.332776°d | 36° | 84° | | | 0° |
| IIa | Thomas et al., 1997 (HST 1994) | 308° ± 10° | 48° ± 10° | 287° + 1617.332776°d | 22° | 54° | 0° | 145° | 330° |
| IIb | Thomas et al., 1997 (HST 1994 + 1996) | 301° ± 5° | 41° ± 5° | 292° + 1617.332776°d | 27° | 38° | 0° | 145° | 330° |
| III | Drummond et al., 1998 (uses pole from IIb) | 301° ± 5° | 41° ± 5° | 310° + 1617.332485°d | 27° | 38° | 350° | 135° | 0° |
| IV | Drummond & Christou, 2008 | 306° ± 7° | 37° ± 7° | 304° + 1617.332485°d | 32° | 44° | 30° | 175° | 0° |
| V | Li et al., 2011 | 305.8° ± 3.1° | 41.4° ± 1.5° | 285.8° + 1617.332776°d | 27.6° | 45.6° | 0° | 145° | 330° |
| VI | Archinal et al., 2011 (uses pole from V) | 305.8° ± 3.1° | 41.4° ± 1.5° | 292° + 1617.332776°d | 27.6° | 45.6 | 354° | 139° | 324° |
| VII | Russell et al., 2012 | 309.03° ± 0.01° | 42.23° ± 0.01° | 75.39° + 1617.333122°d | 27.46° | 50.91° | 210° | 356° | 180° |

1013  **Table 4.** The observing geometries of Dawn FC data and HST 2007 data we used to
1014  compare the lightcurves of Vesta. The last column lists the difference between sub-solar
1015  longitude and sub-observer longitude. A positive value indicates that the morning side is
1016  imaged, while a negative value indicates that the afternoon side is imaged.

1017

| Data Set | Date | Sub-Solar Latitude | Sub-observer latitude | Phase angle | Lat_sun – Lat_obs |
|---|---|---|---|---|---|
| Dawn FC | 2011-06-30 | -25.7° | -32.2° | 26.1° | 29.1° |
| HST 2007 | 2007-05-14/16 | -5.3° | -13.4° | 9.5° | -5.4° |

1018

1019  **Table 5.** List of albedo and spectral units identified in HST maps of Vesta and their
1020  corresponding IAU approved names from Dawn FC data. A complete list of named
1021  features on Vesta from the Dawn mission is available at
1022  http://planetarynames.wr.usgs.gov/Page/VESTA/target
1023
1024

| HST Albedo/Spectral Unit | IAU Approved Feature from Dawn FC Map | Type of Feature | Coordinates (Claudia System) | |
|---|---|---|---|---|
| | | | Latitude | Longitude |
| "Olbers" | Dark ejecta E of Marcia, Calpurnia, Minucia | Albedo unit | -10° to +20° | 197° to 215° |
| #13 | Eusebia | Crater | -42.2° | 204.2° |
| #11 | Vibidia | Crater | -26.9° | 220.1° |
| Q | Numisia | Crater | -7.0° | 247° |
| Z | Teia | Crater | -3.4° | 271° |
| #15 | Oppia | Crater | -8° | 309° |
| #3 | Canuleia, Justinia area | Craters | -33.7° to -34.4° | 294.5° to 317.9 |
| Y | Feralia | Planitia | +4° | 312° |
| K | Marcia | Crater | +10° | 190° |
| A | Octavia ejecta | Ejecta | | |
| B | Octavia | Crater | -3.3° | 147° |
| #8 | Lucaria Tholus | Hill | -13° | 104° |
| #6 | Pinaria area | Crater | -29° | 32° |
| #4 | Rubria, Occia area | Craters | -7.4° to -15.4° | 18.4° |
| #5 | Aquilia | Crater | -49.7° | 41° |
| #7 | Matronalia Rupes | Scarp | -49.5° | 82.7° |

1025
1026

**Figure Captions**

**Figure 1.** Location of Claudia crater in Vesta's western hemisphere with a close up (b) of Claudia's location southeast of merged craters and a detailed view of Claudia crater itself (c).

**Figure 2.** Rotational lightcurves as observed by Dawn FC during approach to Vesta and by HST in May 2007. The observing geometries are listed in Table 3. In all three panels, red and orange symbols are measured from Dawn FC images, and green symbols are from HST images. Dawn FC observations and HST observations are arbitrarily shifted to align with each other in magnitude at each wavelength. The top panel shows lightcurves from Dawn FC clear filter (F1, effective wavelength 699 nm) in red and F7 (652 nm) in orange, and from HST WFPC F673N filter (673 nm); middle panel from Dawn FC F8 filter (438 nm) and HST F439W filter (431 nm); and lower panel Dawn FC F5 filter (961 nm) and HST F953N filter (953 nm). All lightcurves are plotted with respect to sub-solar longitude in IAU 'Olbers' coordinate system.

**Figure 3.** (A) Global topography of Vesta from Hubble imaging data (Thomas, et al, 1997) and from Dawn Framing Camera (B) in IAU 'Olbers' coordinate system. Although the absolute scale ranges from -12 to +12 km in HST data and -22 to +19 km in Dawn data, Vesta's relative topography from HST is consistent with Dawn.

1049 **Figure 4.** (A) HST map of Vesta in 0.673-μm filter projected in the Thomas et al.
1050 (1997a) coordinate system based on observations from 1994, 1996 and 2007 oppositions
1051 at a resolution of ~50 km/pixel. (B) Dawn FC map of Vesta in 0.75-micron filter from
1052 RC1 at a resolution of 9.06 km/pixel with prime meridian similar to Thomas et al.
1053 (1997a) coordinate system in Fig. 4a. The pole position ($\alpha_0$ = 309.03° ± 0.01°, 42.23° ±
1054 0.01°) is the updated one from Russell et al. (2012) and hence features in the Dawn FC
1055 map are slightly rotated with respect to HST map. Fig. 4c is similar to Fig. 4b but the
1056 prime meridian is defined in the Claudia coordinate system (Russell et al. 2012) used by
1057 the Dawn science team in all their publications.

1058

1059 **Figure 5.** Band depth ratio (0.75/0.92 μm) map of Vesta from HST (A) and Dawn (B)
1060 showing the intensity of the pyroxene 0.90-μm absorption feature in the Thomas et al.
1061 (1997a) coordinate system. Areas in red have deeper pyroxene band and areas in blue
1062 have weaker bands. HST and Dawn band depth maps are consistent with each other.

1063

1064 **Figure 6.** Dawn FC color maps of Vesta using data obtained during the RC3B approach
1065 phase of the mission in the Thomas et al. (1997a) coordinate system. (A) Albedo map in
1066 0.75 μm filter, (B) Band depth (0.75/0.92 μm) ratio map with red areas depicting deeper
1067 0.90-μm pyroxene band, and (C) eucrite-diogenite ratio (0.98/0.92 μm) map with red
1068 areas indicating more diogenitic material and blue areas more eucritic.

1069

1070 **Figure 7.** Albedo map of Vesta from HST (A) and Dawn (B) showing corresponding
1071 bright features and dark features identified by Binzel et al. (1997) and Li et al. (2010).

1072  Both maps are in the Thomas et al. (1997a) coordinate system. A global map of Vesta

1073  obtained by the Dawn FC camera in Claudia coordinate system is available online at

1074  http://planetarynames.wr.usgs.gov/images/vesta.pdf

1075

1076  **Figure 8.** Comparison of compositional units observed in HST map of Vesta from Li et

1077  al. (2010) and band depth ratio map from Dawn. Both maps are in the Thomas et al.

1078  (1997a) coordinate system.

1079

1080  **Figure 9.** Comparison of compositional units observed in ground-based map of Vesta

1081  from Gaffey (1997) modified by Binzel et al. (1997) and band depth ratio map from

1082  Dawn. Both maps are in the Thomas et al. (1997a) coordinate system. Note that this map

1083  originally presented in both Gaffey (1997) and modified by Binzel et al. (1997) has

1084  wrong latitude due to incorrect calculation of sub-Earth latitude. While the longitudinal

1085  location of the compositional units in ground-based maps is well constrained (limited by

1086  lightcurve resolution), latitudinal location is weakly constrained as the maps were created

1087  using disk integrated data. Due to weak constrains on the latitudinal location of

1088  compositional units in Gaffey (1997) and Binzel et al. (1997) maps, comparison with

1089  Dawn maps remains valid despite the latitude error.

1090

1091

1092

1093

1094

Figure 1. Dawn, HST, Ground-based Studies of Vesta

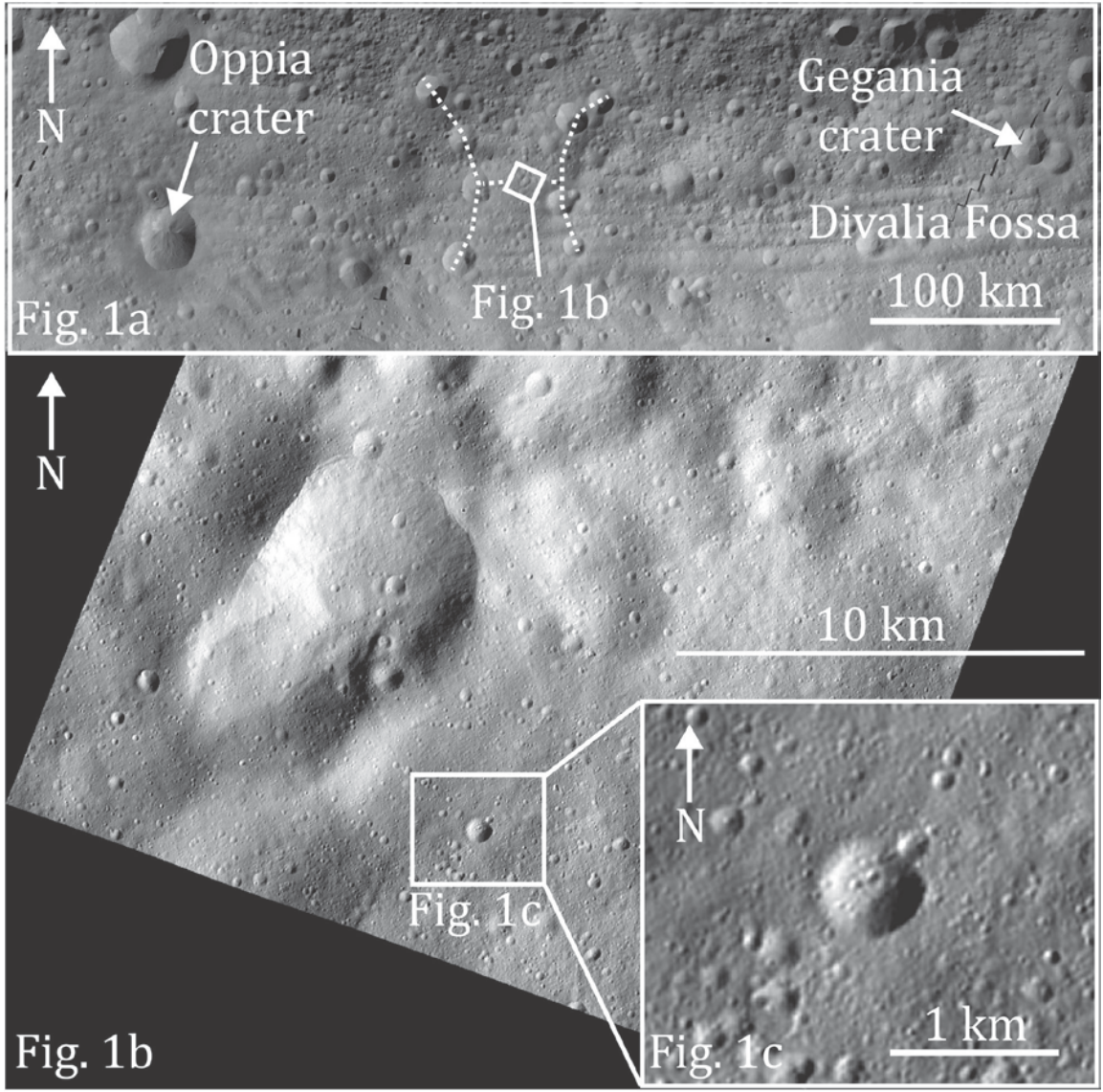

1103    Figure 2. Dawn, HST, Ground-based Studies of Vesta

1104

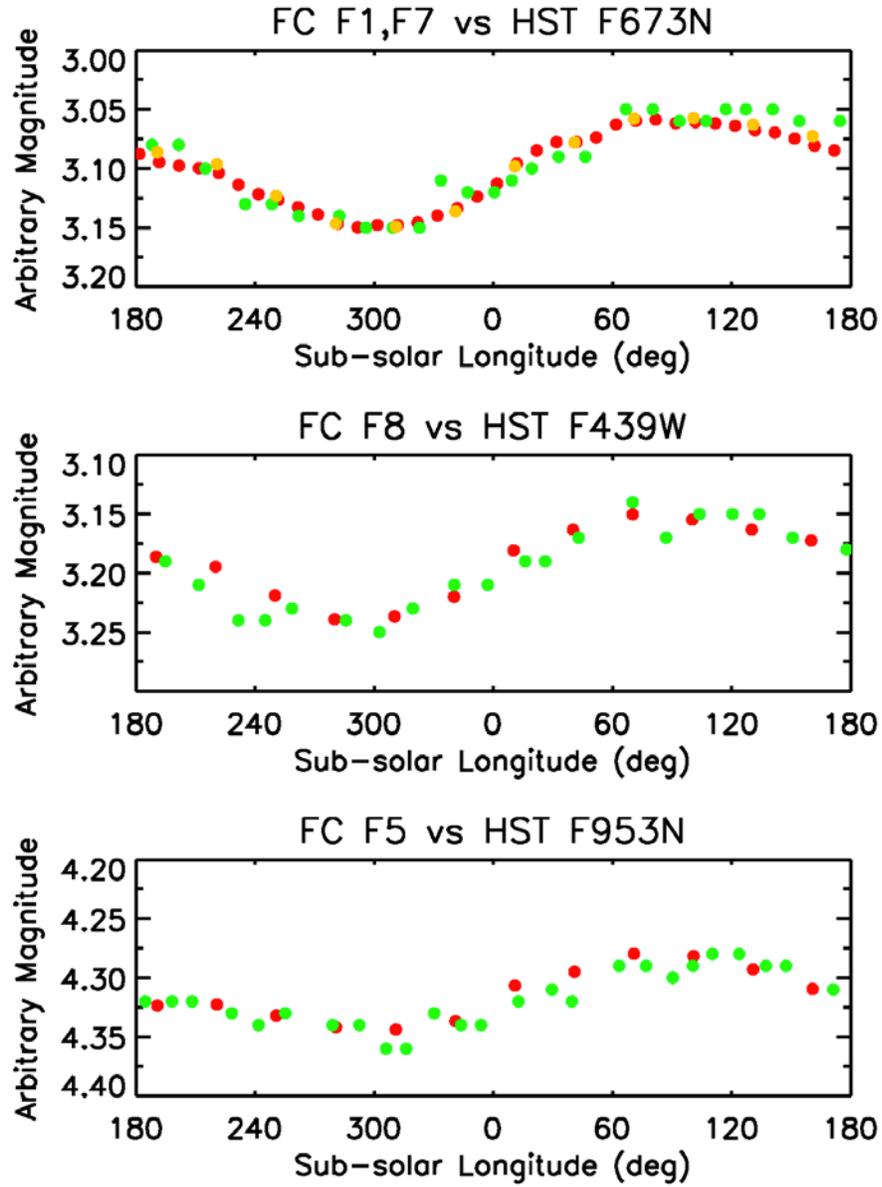

1105

1106

1107    Figure 3. Dawn, HST, Ground-based Studies of Vesta

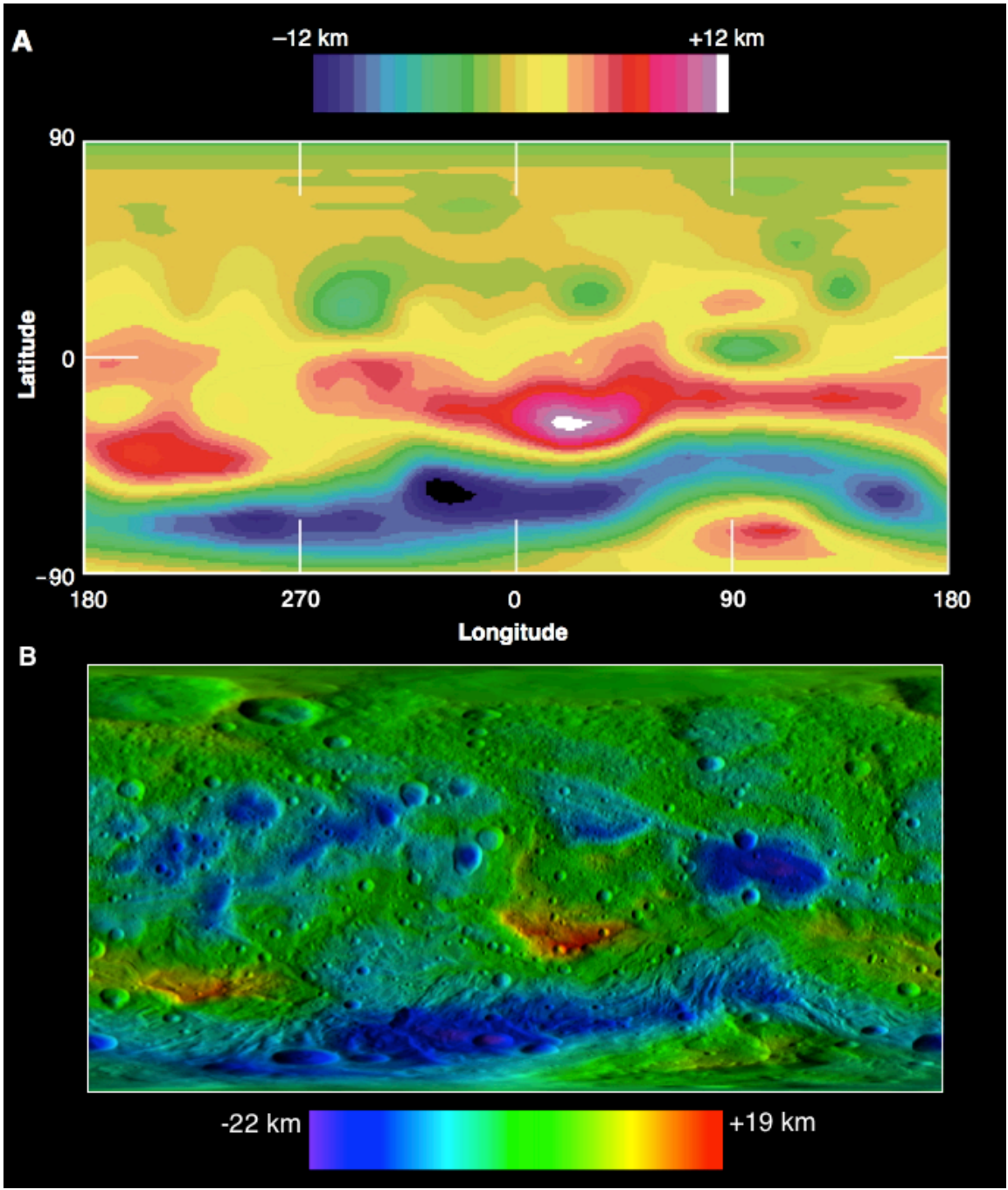

1112    Figure 4. Dawn, HST, Ground-based Studies of Vesta

1113

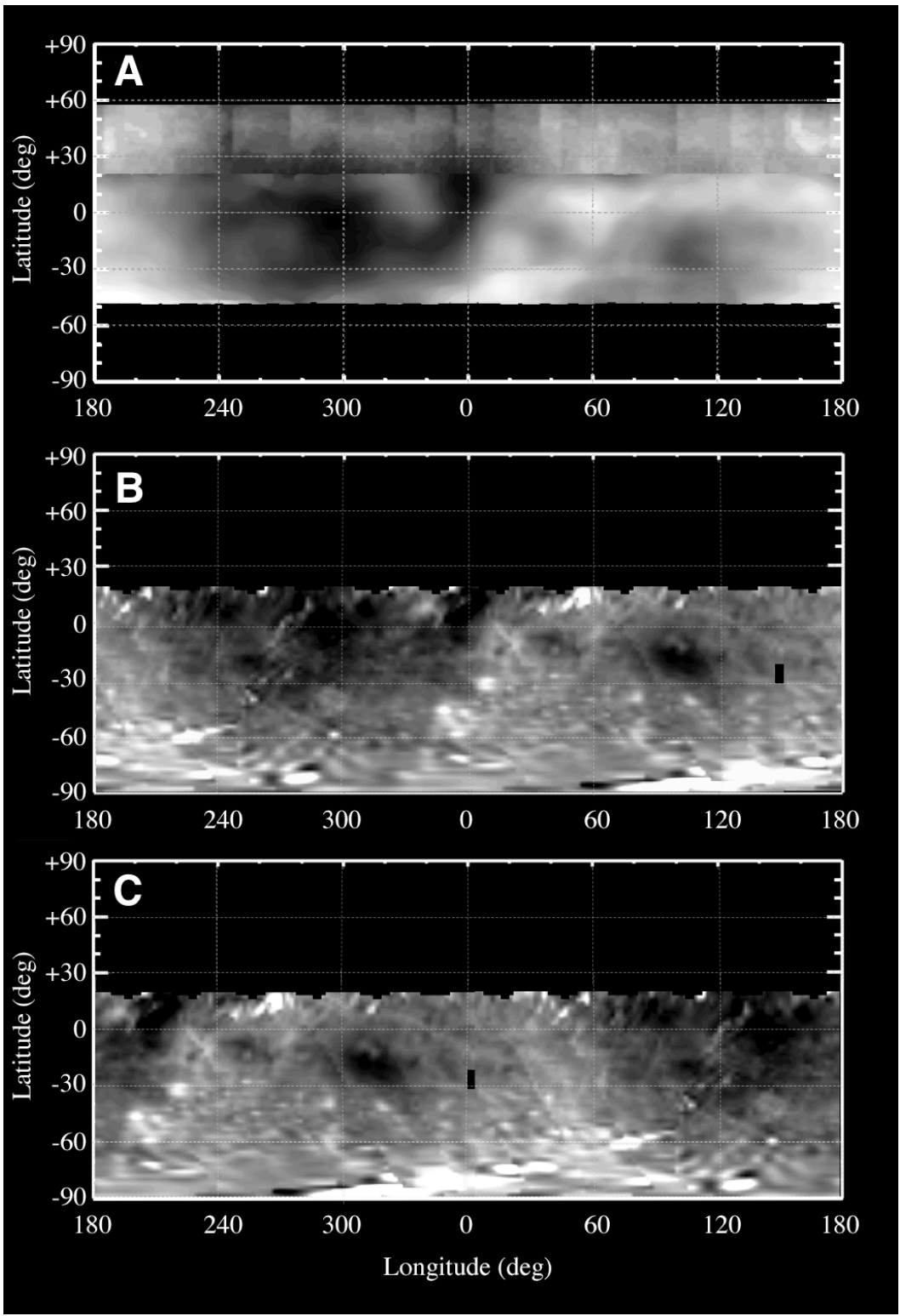

1114

Figure 5. Dawn, HST, Ground-based Studies of Vesta

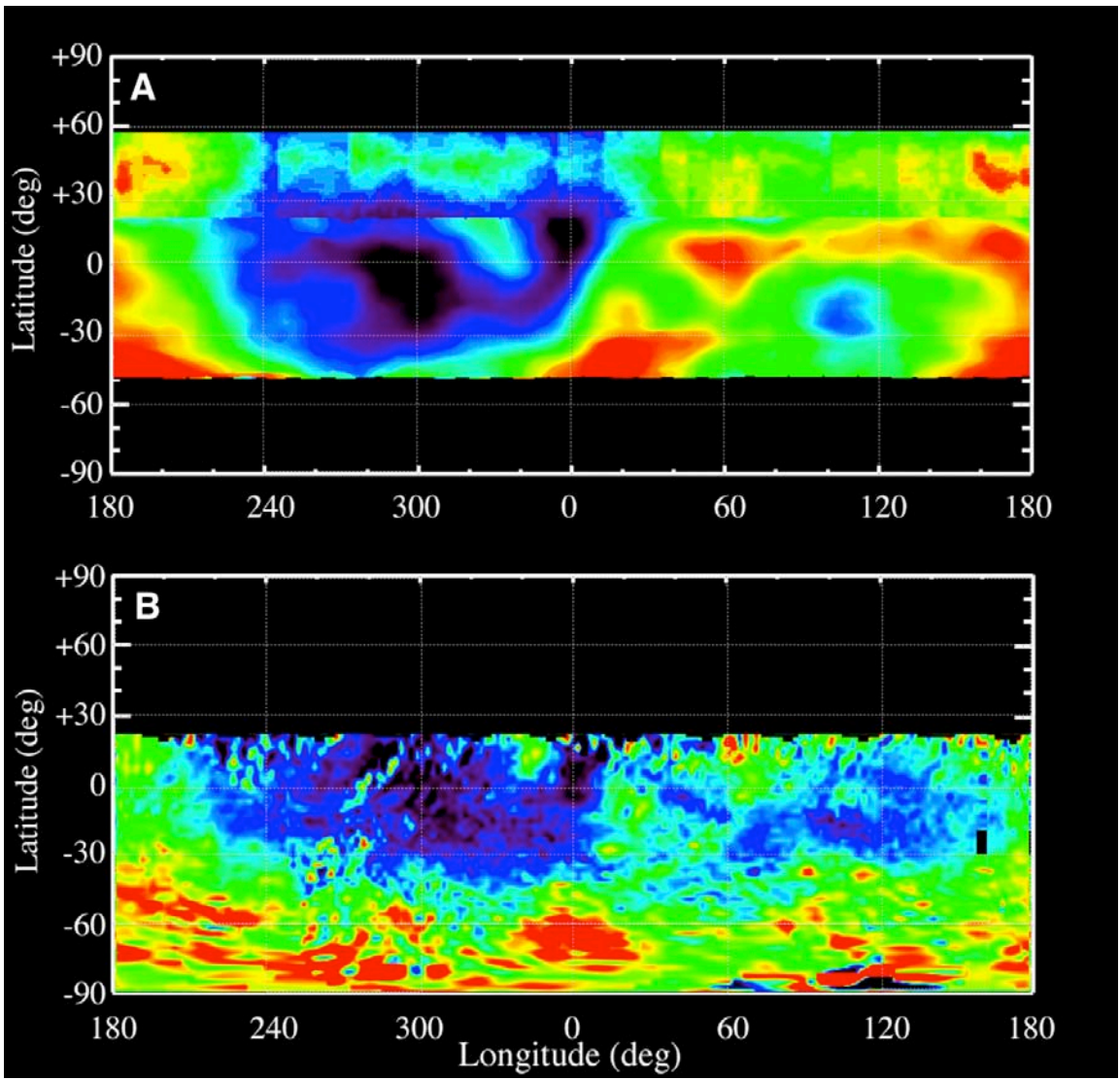

1124 Figure 6. Dawn, HST, Ground-based Studies of Vesta

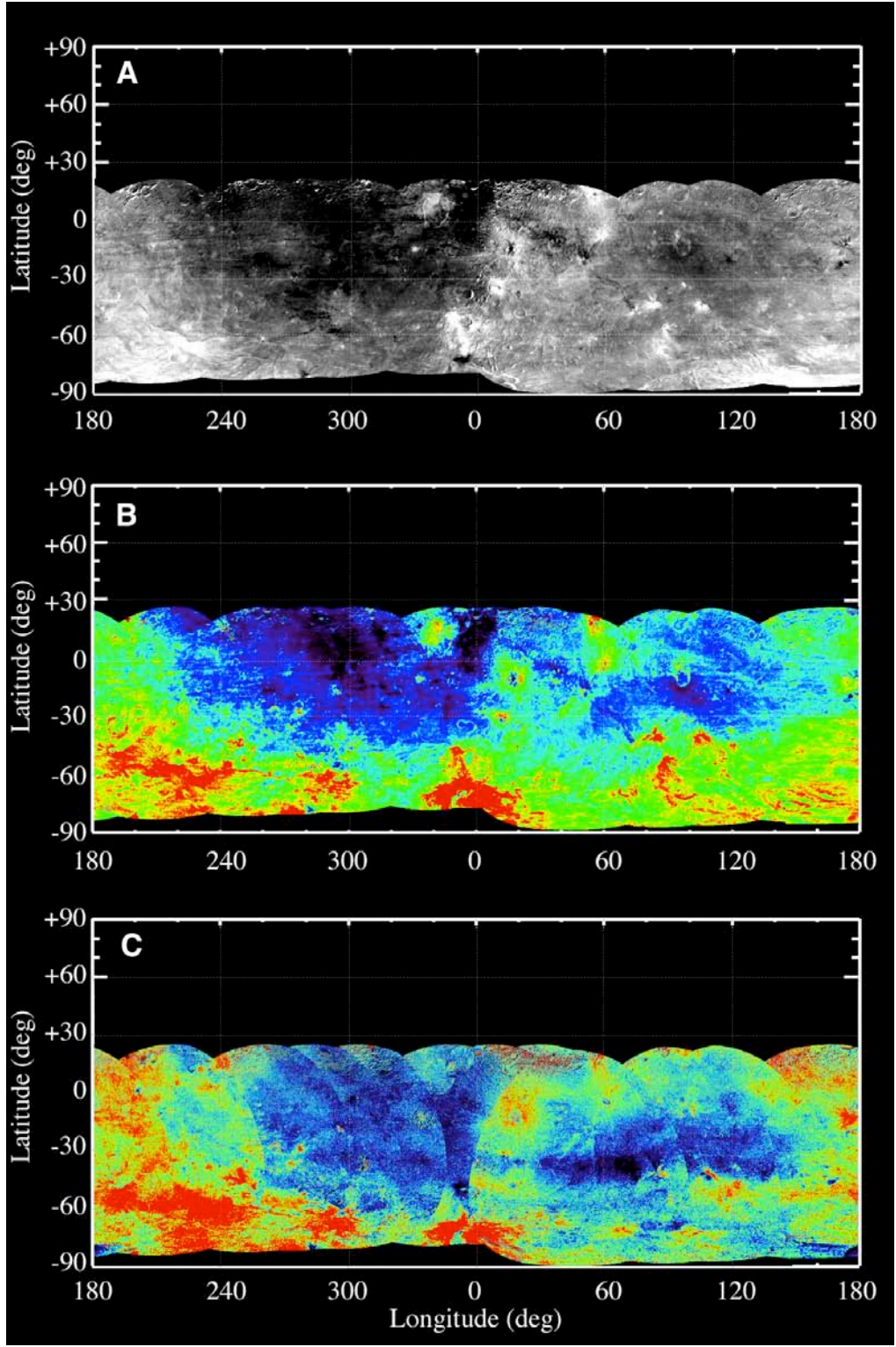

1125

1126

1127    Figure 7. Dawn, HST, Ground-based Studies of Vesta

1128

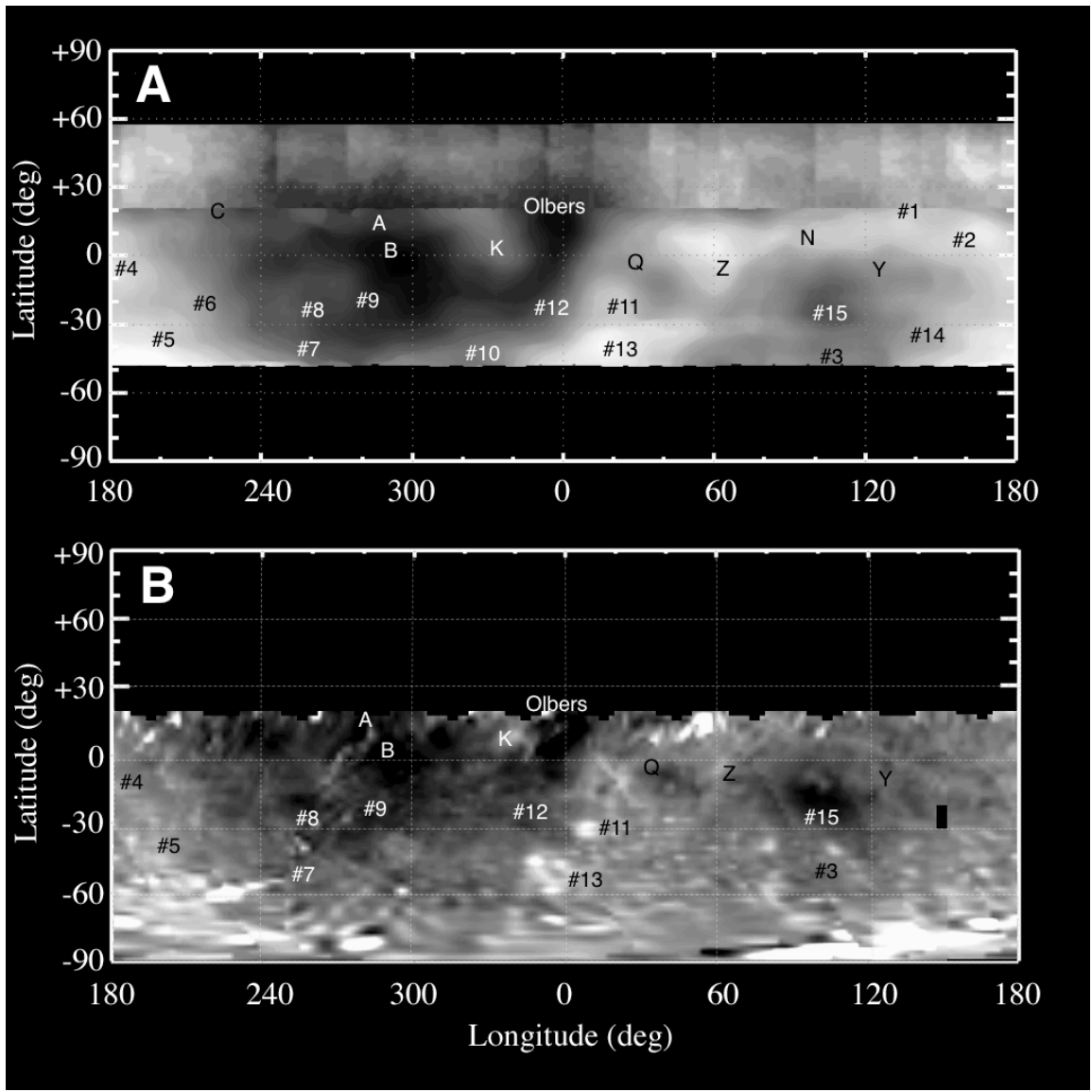

1129
1130
1131
1132
1133
1134

1135    Figure 8. Dawn, HST, Ground-based Studies of Vesta

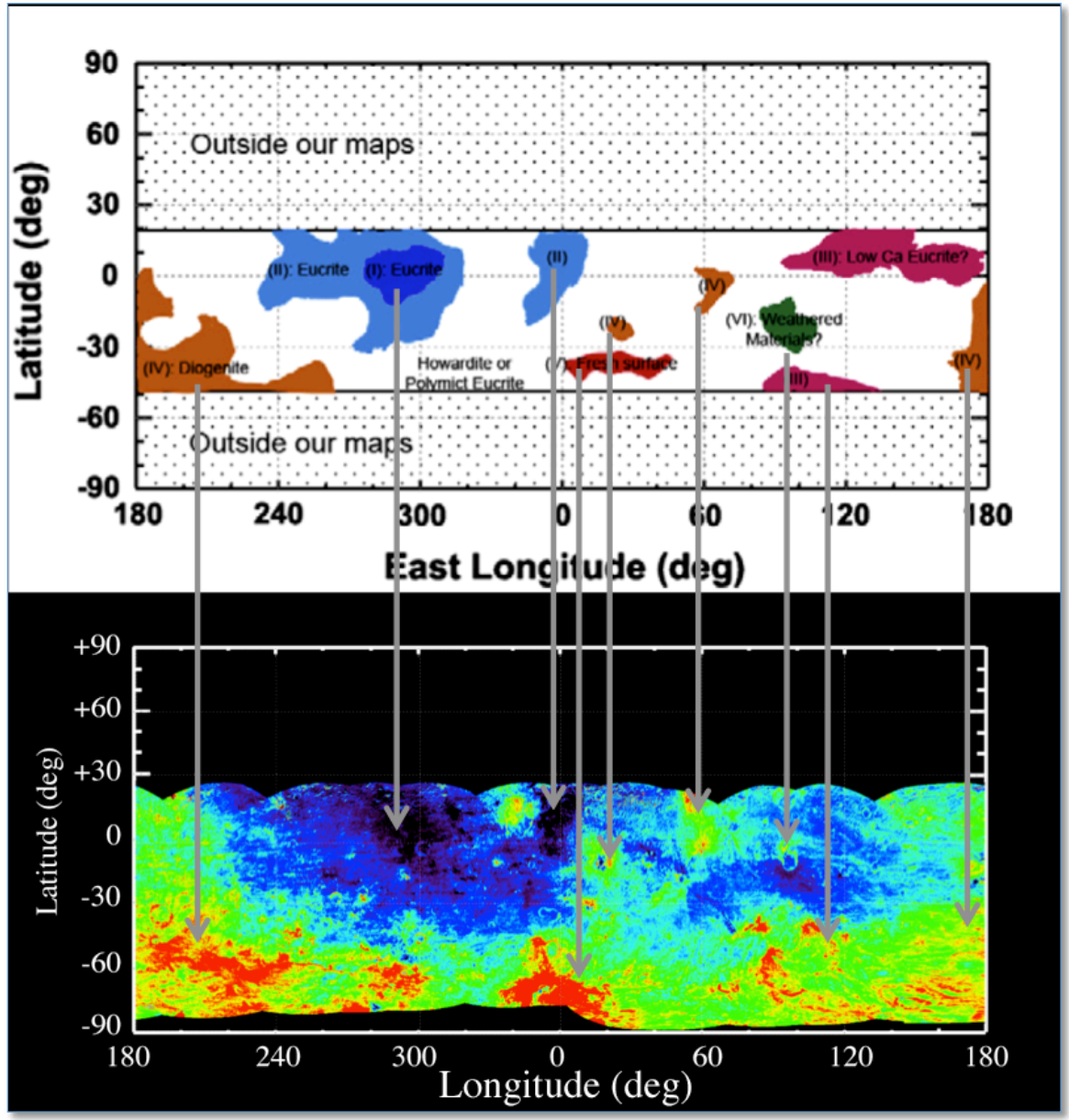

1142    Figure 9. Dawn, HST, Ground-based Studies of Vesta

1143

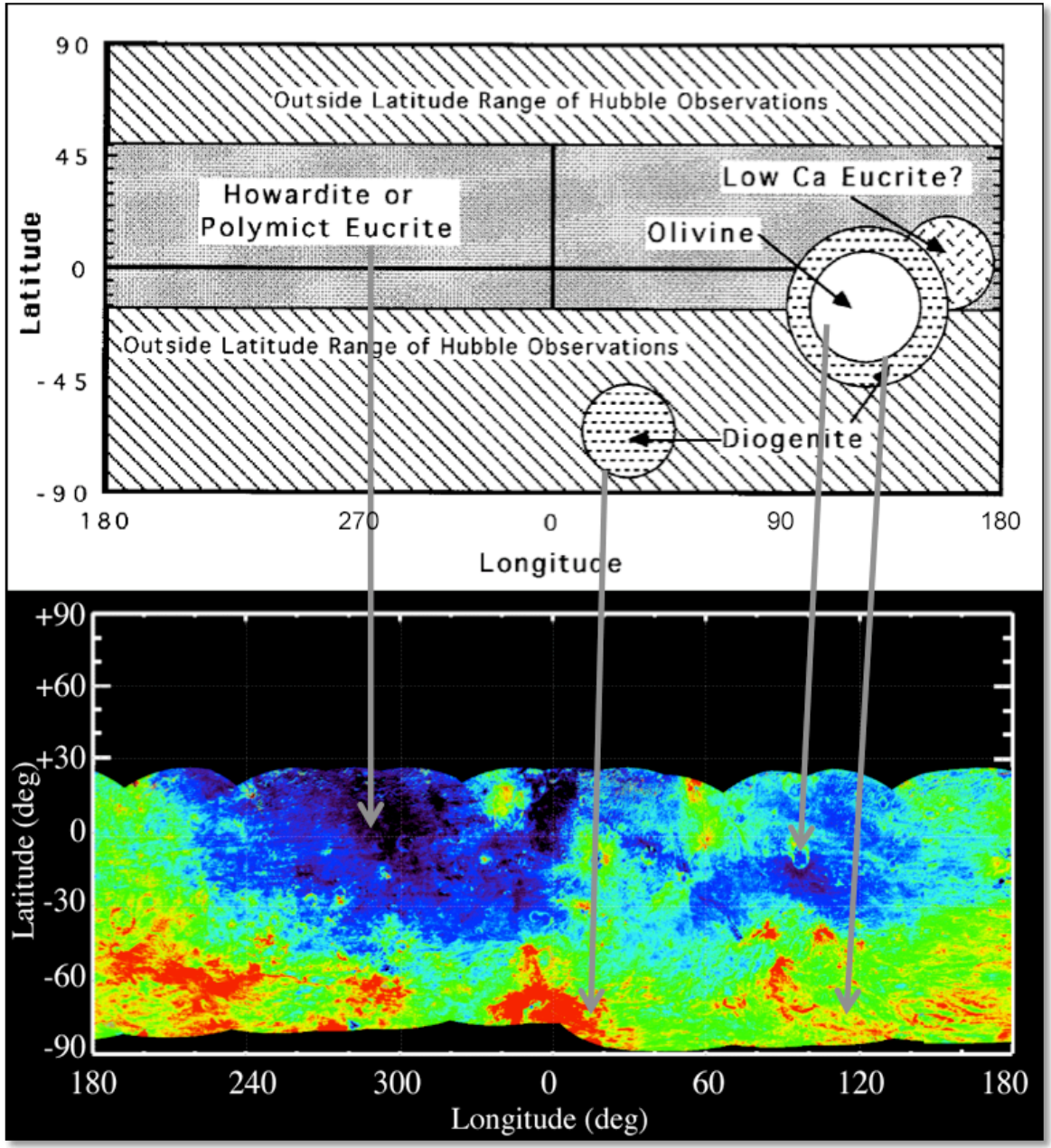

1144
1145
1146
1147